\documentclass[a4paper,fleqn,usenatbib]{mnras}

\usepackage{newtxtext,newtxmath}
\usepackage[T1]{fontenc}
\usepackage{ae,aecompl}

\usepackage{graphicx}	% Including figure files
\usepackage{amsmath}	% Advanced maths commands
\usepackage{amssymb}	% Extra maths symbols
\usepackage{graphicx}	% Including figure files
\usepackage{amsmath}	% Advanced maths commands
\usepackage{amssymb}	% Extra maths symbols
\usepackage{rotating}
\usepackage{lscape}
\usepackage{comment}
\usepackage[utf8]{inputenc}
\usepackage[dvipsnames]{xcolor}
\usepackage{siunitx}
\usepackage{verbatim}
\usepackage[english]{babel}
\usepackage[T1]{fontenc}
\usepackage{ae,aecompl}
\usepackage{rotating}
\usepackage{textcomp}
\usepackage[normalem]{ulem}
\usepackage{soul}

\title[Neural network techniques]{Optimizing neural network techniques in classifying {\it Fermi}-LAT gamma-ray sources}

\author[M. Kova{\v{c}}evi{\'{c}} et al.]
{
M. Kova{\v{c}}evi{\'{c}}$^{1}$\thanks{E-mail: milosh.kovacevic@gmail.com},
G. Chiaro$^{2}$\thanks{E-mail: graziano.chiaro@inaf.it},
S. Cutini$^{1}$,
G. Tosti$^{3}$
\\
\\
$^{1}$INFN -- Istituto Nazionale di Fisica Nucleare Sez. Perugia, I-06123 Perugia, Italy\\
$^{2}$INAF -- Istituto di Astrofisica Spaziale e Fisica Cosmica - Milano, I-20133 Milano, Italy\\
$^{3}$Dipartimento di Fisica e Geologia, Univ. degli Studi di Perugia, I-06123 Perugia, Italy\\
}

\date{Accepted 2019 October 14, Received October 10; in original form 2019 May 27}

\pubyear{2019}

\begin{document}
\label{firstpage}
\pagerange{\pageref{firstpage}--\pageref{lastpage}}
\maketitle

\begin{abstract}  
{Machine learning is an automatic technique that is revolutionizing scientific research, with innovative applications and wide use in astrophysics. The aim of this study was to developed an optimized version of an Artificial Neural Network machine learning method for classifying blazar candidates of uncertain type detected by the {\it Fermi} Large Area Telescope (LAT) $\gamma$–ray instrument. The initial study used information from $\gamma$-ray light curves present in the LAT 4-year Source Catalog. In this study we used additionally $\gamma$-ray spectra and multiwavelength data, and certain statistical methods in order to improve classification. The final result of this study increased the classification performance by about 80$\%$ with respect to previous method, leaving only 15 unclassified blazars instead of 77 out of total 573 in the LAT catalog. Other blazars were classified into BL Lacs and FSRQ in ratio of about two to one, similar to previous study. In both studies a precision value of 90\% was used as a threshold for classification.}
\end{abstract}

\begin{keywords}
methods: statistical -- galaxies: active -- gamma-rays: galaxies -- BL Lacertae objects: general.
\end{keywords}

\section{Introduction}

Since August 2008 the {\it Fermi} Large Area Telescope (LAT) provides the most comprehensive view of the $\gamma$-ray sky in the 100 MeV to 300 GeV energy range \citep{fermi}. The LAT 4-year  Source Catalog \textit{3FGL} \citep{3fgl} listed 3033 $\gamma$-ray sources  of which 1717 were blazars, including 573 blazar candidates of uncertain type (BCU). In addition 1010 of the detected sources had not even a tentative association with a likely $\gamma$-ray emitting source. As a result, the nature of about half the $\gamma$-ray sources is still not completely known even if, because blazars are the most numerous $\gamma$-ray source class, it could be reasonable to expect that a large fraction of unassociated sources might belong to the BL Lacertae (BL Lac) or Flat Spectrum Radio Quasar (FSRQ) class. When rigorous classification analyses are not available, machine learning techniques (MLTs) represent powerful tools that enable identification of uncertain objects based on their expected classification. Machine learning is a data analytics technique that teaches computers to do what comes naturally to humans and animals: learn from experience. Traditional computer programs do not consider the output of their tasks, and therefore they are unable to improve their efficiency. MLT addresses this exact problems and involves the creation of an algorithm that is able to learn and therefore improve its performances by gathering more data and experience. MLT uses identified objects to teach the algorithm to distinguish each source class on the basis of parameters that describe its intrinsic features. The algorithm adaptively improves its performance as the number of samples available for learning increases.
The algorithm, under certain conditions, generates an output that can be interpreted as a Bayesian a posteriori probability modeling the likelihood of membership class on the basis of input parameters \citep{gis90, ric91}. In  this work we explore the possibility to improve the performance of a machine learning algorithm  \citet{bflap} based on the variability of blazars, applying new physical parameters that characterize  the nature of those sources and some statistical adjustments in order to increase the accuracy of the algorithm, making it more efficient and effective.

The expected result should be an optimized algorithm that is able to estimate, with more precision than in the past, the number of uncertain blazars that could belong to the BL Lac or FSRQ class in the {\it Fermi}-LAT Source Catalogs.

The paper is organized as follows: in Section ~\ref{<2>} we provide a brief description of the main features of the most frequently used machine learning techniques in astrophysics. In Section ~\ref{<3>} we present our optimization of an Artificial Neural Network (ANN) method. In Section ~\ref{<4>} we present the results and compare the performance of the optimized algorithm against the original one. We discuss predictions and implication of our results in Section ~\ref{<5>}.

\section {Machine learning techniques}
\label{<2>}

\begin{figure}
\begin{center}
\includegraphics[width=.35\textwidth]{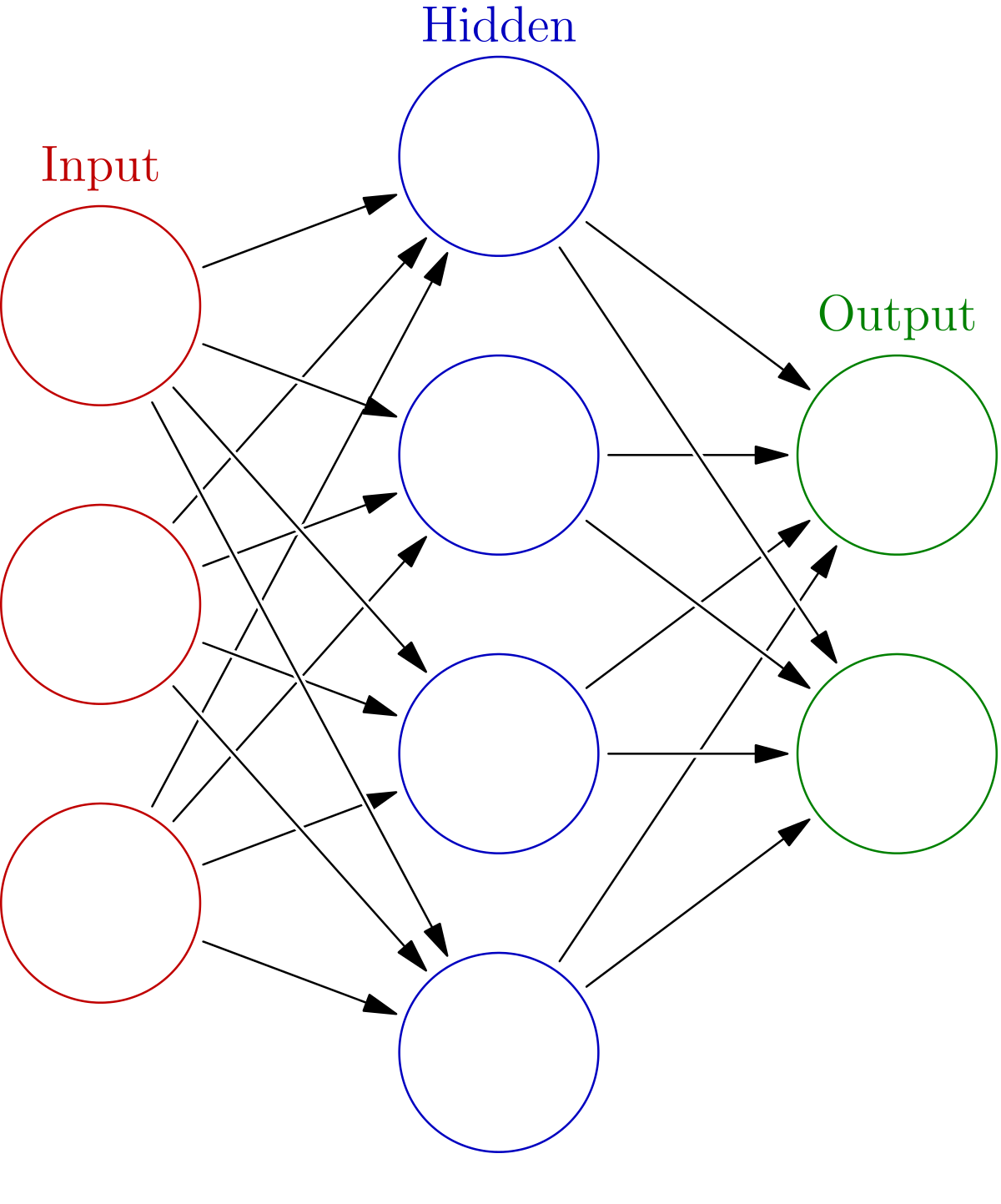}
\caption{Schematic view of a Two Layer Perceptron (2LP), the Artificial Neural Network architecture. Each circle represents a single neuron. Each arrow represents association between output values of neurons to the weights of neurons in the successive layer. Outputs of neurons in the \textit{Input layer} are just values of parameters describing an astronomical source. Data enter the 2LP through the nodes in the input layer. The information travels from left to right across the links and is processed in the nodes through an activation function. Each node in the output layer returns the likelihood of a source to belong to the specific class. 
\label{<o>}}
\end{center}	
\end{figure}

In previous studies, \citet{ack2012, lee2012, hassan, doert2014, bflap, mirabal, pablo, lefau, zoo} and other authors, have explored the application of MLT classifying undetermined  $\gamma$-ray sources in {\it Fermi}-LAT $\gamma$-ray source catalogs. The first study was applied to the 1-year  Source Catalog \textit{1FGL} \citep{1fgl}, the next 3 studies to the 2-year Source Catalog \textit{2FGL} \citep{2fgl} and the rest were applied to the 4-year  Source Catalog \textit{3FGL} \citep{3fgl}. Here we briefly introduce the general features of the most frequently used MLTs  in astrophysics  for such cases. 
\begin{itemize}

\item{
{\bf{The Random Forest.}} 
The Random Forest method (RF) \citep{rf, lia} is an ensemble learning method that uses decision trees as building blocks for classification, regression and other tasks. By aggregating the predictions based on a large number of decision trees, RF generally improves the overall predictive performance while reducing the natural tendency of standard decision trees to overfit the training set. The RF package also computes the proximity measure, which, for each pair of elements (i, j), represents the fraction of trees in which elements i and j fall in the same terminal node. This can be used to calculate the {\it{outlyingness}} of a source, as the reciprocal of the sum of squared proximities between that source and all other sources in the same class, normalized by subtracting the median and dividing by the median absolute deviation, within each class. \citet{doert2014, hassan, pablo, mirabal} used the RF algorithms in order to classify unassociated sources and uncertain active galactic nuclei (AGNs) from the Fermi $\gamma$-ray source catalogs.
}

\item{
{\bf{The Support Vector Machines.}} 
The Support Vector Machine (SVM) \citep{cor, vap} is a discriminative classifier formally defined by a separating hyperplane. In other words, given labeled training data (supervised learning), the algorithm outputs an optimal hyperplane which categorizes new examples. In two-dimensional space this hyperplane is a line dividing a plane in two parts where  each class lies on either side. The method maximises the separation between different classes, which can then be used in classification or regression analysis. In \citet{hassan} the authors used a SVM algorithm and the Random Forest algorithm building a classifier that can distinguish between two AGN classes: BL Lac and FSRQ based on observed $\gamma$-ray spectral properties. Combining both methods they managed to classify 235 out of 269 uncertain AGNs from the 2FGL catalog into BL Lacs and FSRQs with 85\% accuracy.
}

\item{
{\bf{The Boosted Decision Trees.}} 
The Boosted Decision Tree (BDT) \citep{fre} is based on the decision trees, a classifier structured on repeated {\it{yes/no }} decisions designed to separate {\it{positive}} and {\it{negative}} classes of events. Thereby, the phase space of the discriminant parameters is split into two different regions and generates a forest of weak decision trees and combines them to provide a final strong decision. At each step, misclassified events are given an increasing weight. \citet{lefau} used BDT together with ANN in order to classify unassociated sources and uncertain blazars (BCU) in the 3FGL catalog. Selecting 486 unflagged BCUs, the authors classified 295 of them as BL Lacs with 13 predicted false associations, and 146 as FSRQs with 39 predicted false associations. Both MLT methods were found to perform similarly.
}

\item{
{\bf{The Artificial Neural Network.}}\\
The Artificial Neural Network (ANN) \citep{ann} is probably the most used machine learning technique in astrophysics. Regarding {\it Fermi}-LAT sources, ANN algorithms were used in \citet{bflap} and \citet{zoo} for classifying uncertain blazars and were also used in the above-mentioned work of \citet{doert2014} and \citet{lefau} for classifying unassociated sources and uncertain blazars.

Basic units of neural networks are \textit{neurons} which are organized into \textit{layers} and are connected to each other. Neurons, layers and lines connecting them are abstract mathematical concepts which help to visualise how the input values (describing an astronomical source in our case) to the network are transformed in order to obtain classification for that source.

A standard neural network consists of an input layer, one or more hidden layers and an output layer. In Fig.~\ref{<o>} the schematic view of the basic architecture of an ANN algorithm is shown. Neurons in the input layer are just values of input parameters from a single source (flux values in different time bins for example). Each neuron in the first hidden layer has a set of \textit{weights} (numerical values) which are associated to input parameters. The number of weights in each neuron equals the number of input parameters. The association between weights and input parameters is presented by arrows connecting all input neurons to all neurons in the hidden layer. For each neuron in the hidden layer, the sum of products between each weight and input parameter\footnote{A single value \textit{bias} can be added.} is then used in an \textit{activation/transfer function} to create a single output. The outputs of neurons in the hidden layer are then used as input values for all neurons in the successive layer (which is also presented by arrows). Neurons in the output layer produce the final result.

When classification is the goal, the number of neurons in the output layer usually equals the number of classes. The sum of outputs from these neurons (for a single astronomical source) equals 1 and the output value from each neuron is interpreted as the probability of that source belonging to a given class.

Training the network with known/labeled sources involves setting the weights of all neurons in the network so that difference between given outputs and desired outputs (for many sources combined), quantified by a \textit{Loss/Cost function}, is minimized. The sample of sources used in training the network typically contains 50 - 80 \% of all known sources. The rest are divided into two independent samples -- \textit{validation sample} and \textit{test sample}, which are used to avoid overfitting and to evaluate the network on sources it has not seen during training.

Basically, what this means is that ANN is a mathematical function over an $N$-dimensional space, where $N$ is the number of input parameters to the network. Training the network involves adjusting the very large number of ANN parameters (weights) in order to find a function which best separates objects belonging to different classes.

The original ANN algorithm that we considered in this study was used for the first time in \citet{bflap} (hereinafter {\it{C16}}) and subsequently in \citet{zoo} (hereinafter {\it{S17}}). The algorithm compared the $\gamma$-ray light-curve of the source under investigation with a template of classified blazar class light curves, then measured the difference in a proper metric. The authors of both papers used a simple neural model known as Two Layer Perceptron (2LP), rather similar to the method used by \citet{lefau} but with a simpler architecture.

In this work we explored possibilities to improve the efficiency of the original algorithm used in {\it{C16}} and subsequently in {\it{S17}}. Even if the original ANN algorithm was very effective, the number of sources with uncertain classification in {\it{C16}} and {\it{S17}} remained consistent. In {\it{C16}} analyzing 573 BCUs, 77 sources remained with uncertain classification. Also in {\it{S17}} classifying with the same algorithm the AGN-{\it{like}} sources, 103 of 559 sources remained with uncertain blazar classification.

In order to optimize the performance, we decided to use additional parameters describing blazars and test different network architectures.
For input parameters to the network we used $\gamma$-ray parameters present in the the 3FGL catalog\footnote{\url{https://fermi.gsfc.nasa.gov/ssc/data/access/lat/4yr_catalog/}.} and multiwavelength data in the {\it Fermi}-LAT 4-year AGN Catalog\footnote{\url{http://www.ssdc.asi.it/fermi3lac/}.} \textit{3LAC} \citep{3lac} which are publicly available. 

Since we expected to perform the process of training the network many times, we decided to use \textit{TensorFlow}\footnote{\url{https://www.tensorflow.org}. TensorFlow is an open source library for machine learning. It is relatively easy to use, provides details on the process of training and options for different network architectures, and is fast -- network can be trained on an ordinary computer in relatively short time.} \citep{tf} which was implemented in \textit{Python}\footnote{\url{https://www.python.org/}.}.
}

\end{itemize}

\section{The method }
\label{<3>}

\subsection{ Gamma-ray variability }
\label{<variability>}

\begin{figure}
\begin{center}
\includegraphics[width=.5\textwidth]{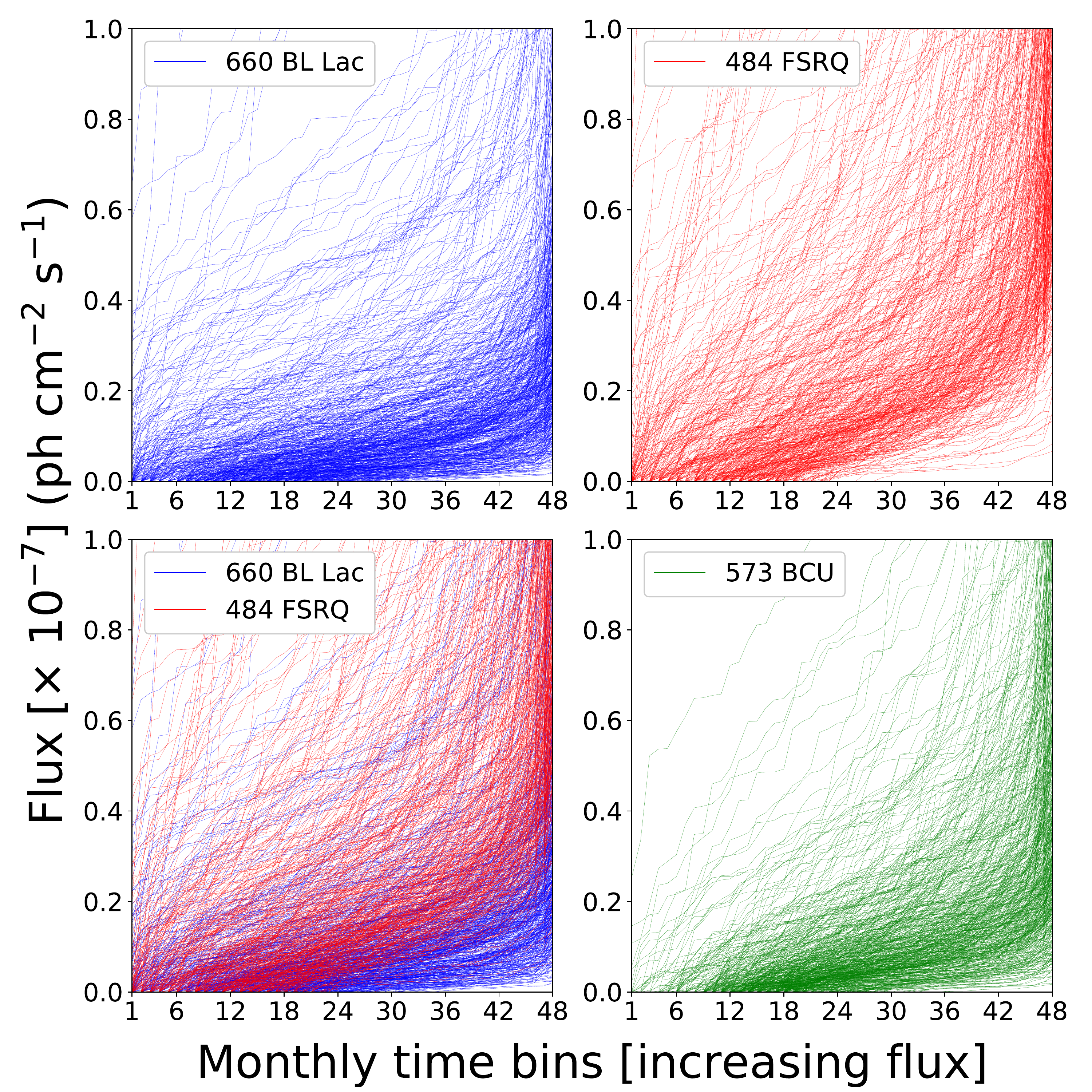}
\caption{Sorted flux curves of 3FGL blazars (4 years of data). Fluxes are in the energy range of 100 MeV - 100 GeV. Each curve represents a single source. Vertical axes present monthly flux values. Horizontal axes present 48 monthly time bins. For each source, the Nth monthly time bin corresponds to the month of observation when the Nth lowest monthly flux was observed. Therefore, lower numbers correspond to months of lower activity for each source while higher numbers to months of higher activity.
BL Lacs are in the top-left, FSRQs in the top-right, both are in the lower-left and BCUs are in the lower-right. Curves for some sources extend beyond the plot limit of 10$^{-7}$ ph cm$^{-2}$ s$^{-1}$. Figure reproduced from {\citet{bflap}.}}
\label{<lcvar>}
\end{center}
\end{figure}

In {\it{C16}} the Empirical Cumulative Distribution Function (ECDF) of the monthly bins of the 3FGL BCU  $\gamma$-ray light curves was applied to the ANN as an estimator able to classify BCUs into BL Lacs and FRSQs. The monthly fluxes are in the energy range of 100 MeV to 100 GeV. In a similar way to ECDFs in {\it{C16}} we obtain curves by sorting monthly flux values from lowest to highest for each source. This produces a set of (12 months $\times$ 4 yr) 48 sorted flux values. These curves contain information on flaring patterns, along with the monthly averaged brightest flares and variability of the sources. The distinctiveness of 3FGL BL Lacs and FSRQs is shown in Fig.~\ref{<lcvar>}. BL Lacs tend to be dimmer than FSRQs. Their emission also tends to be more continuous over time than that of FSRQs, which show more variability. This can be seen in the lower-left plot of Fig.~\ref{<lcvar>}. In the lower-right part of the plot there is an area where mostly BL Lacs are found. Sources passing trough this area are ones which have lower flux ($\lesssim$ 2 $\times$ 10$^{- 8}$ ph cm$^{-2}$ s$^{-1}$) during their brightest months. Both dimmer and brighter BL Lacs tend to have more "horizontal" curves that reflect their lower variability. This result convinced the authors in {\it{C16}} to use the ECDF as the sole ANN parameter to compute the likelihood of their sample of uncertain sources to be BL Lac or FSRQ. Quick comparison {\it{looking by eyes}} of blazar classes in Fig.~\ref{<lcvar>} suggests that BCU ECDFs are closer to BL Lac ones and that the larger part of BCUs could be BL Lacs.

\begin{figure}
\begin{center}
\includegraphics[width=.5\textwidth]{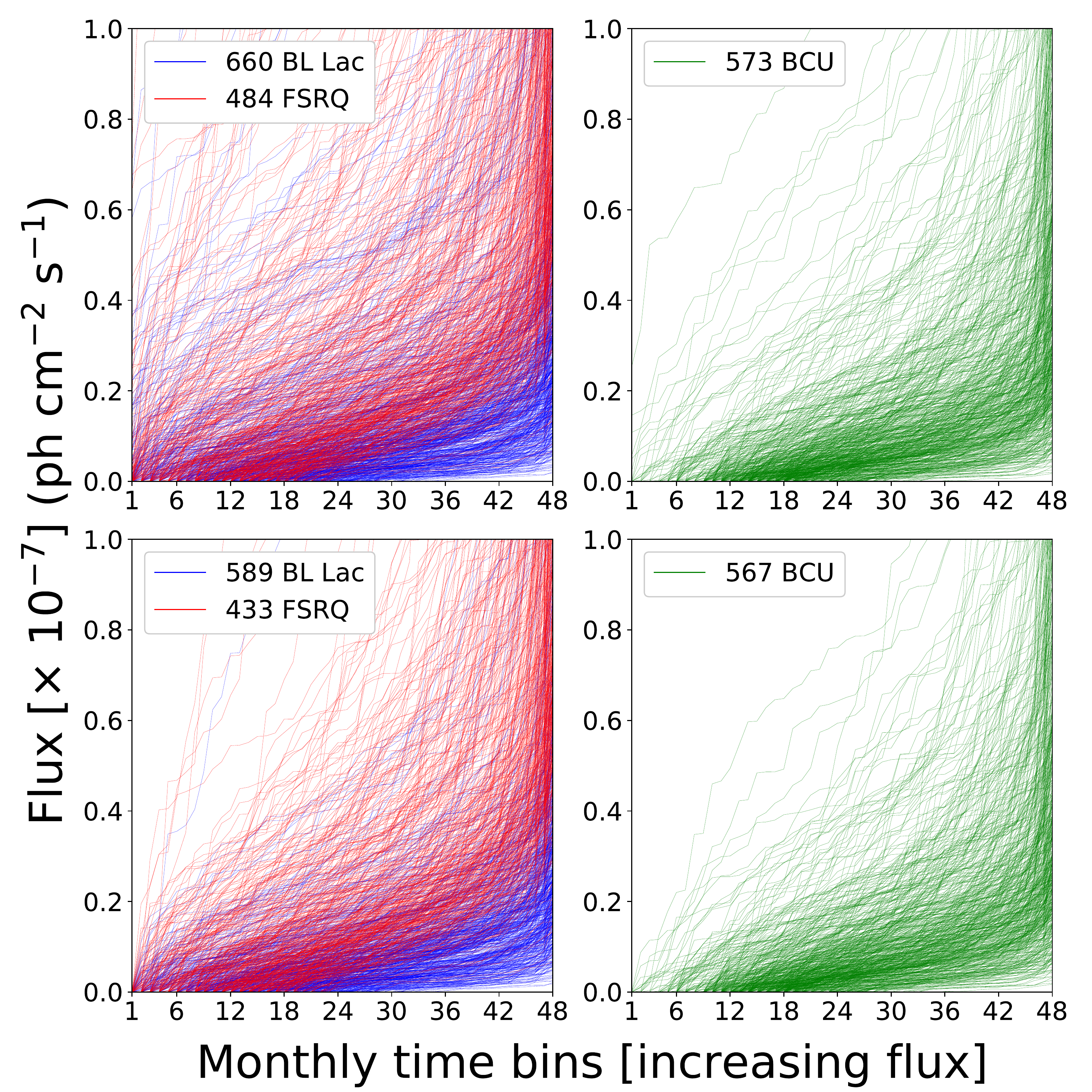}
\caption{The upper two plots contain all the 3FGL blazars while the two plots at the bottom contain 3FGL blazars after applying the flux threshold cut. The number of sources for each class is written on the plots before and after the cut. Blue curves correspond to BL Lacs, red to FSRQs and green to BCUs. 
The sources affected by the cut have flux values above 0 for monthly bin number 1 (the dimmest month). After applying the cut, the upper-left part of the plot for BL Lacs and FSRQs becomes more clear. For BCUs, the same part of the plot remains similar after the cut because there were not many sources passing through it (only 6 sources are removed by the cut).} 
\label{<lcvar cut>}
\end{center}
\end{figure}

Since ECDF curves represent the only set of parameters originally used in {\it{C16}}, it was interesting to test if some statistical methods could improve the final performance of the network. While distinctiveness of BL Lacs and FSRQs is obvious for flux values $\lesssim$ 2 $\times$ 10$^{- 8}$ ph cm$^{-2}$ s$^{-1}$ during the brightest months, they are much more intertwined and similar for sources which have higher flux value during dimmer months (upper-left part of the plots). Since BCUs are hardly present in this part of the plot (parameter space), removing some of these BL Lacs and FSRQs would enable better separation while simultaneously making joint BL Lac and FSRQ distribution more representative of BCU distribution. One way to proceed is to identify sources that have a flux value above a detection threshold for the dimmest month (monthly bin number 1 on the plots in Fig.~\ref{<lcvar cut>}), and then to remove them. Applying this constraint, the number of BL Lacs fell from 660 to 589 (-10$\%$) and FSRQs from 484 to 433 (-10$\%$) . The reduction of source number did not affect training and testing the network. The number of BCUs fell from 573 to 567, so only six sources were lost for classification (Fig.~\ref{<lcvar cut>}).

\subsection{ Gamma-ray spectrum}
\label{<spectrum>}

\begin{figure}
\begin{center}
\includegraphics[width=.5\textwidth]{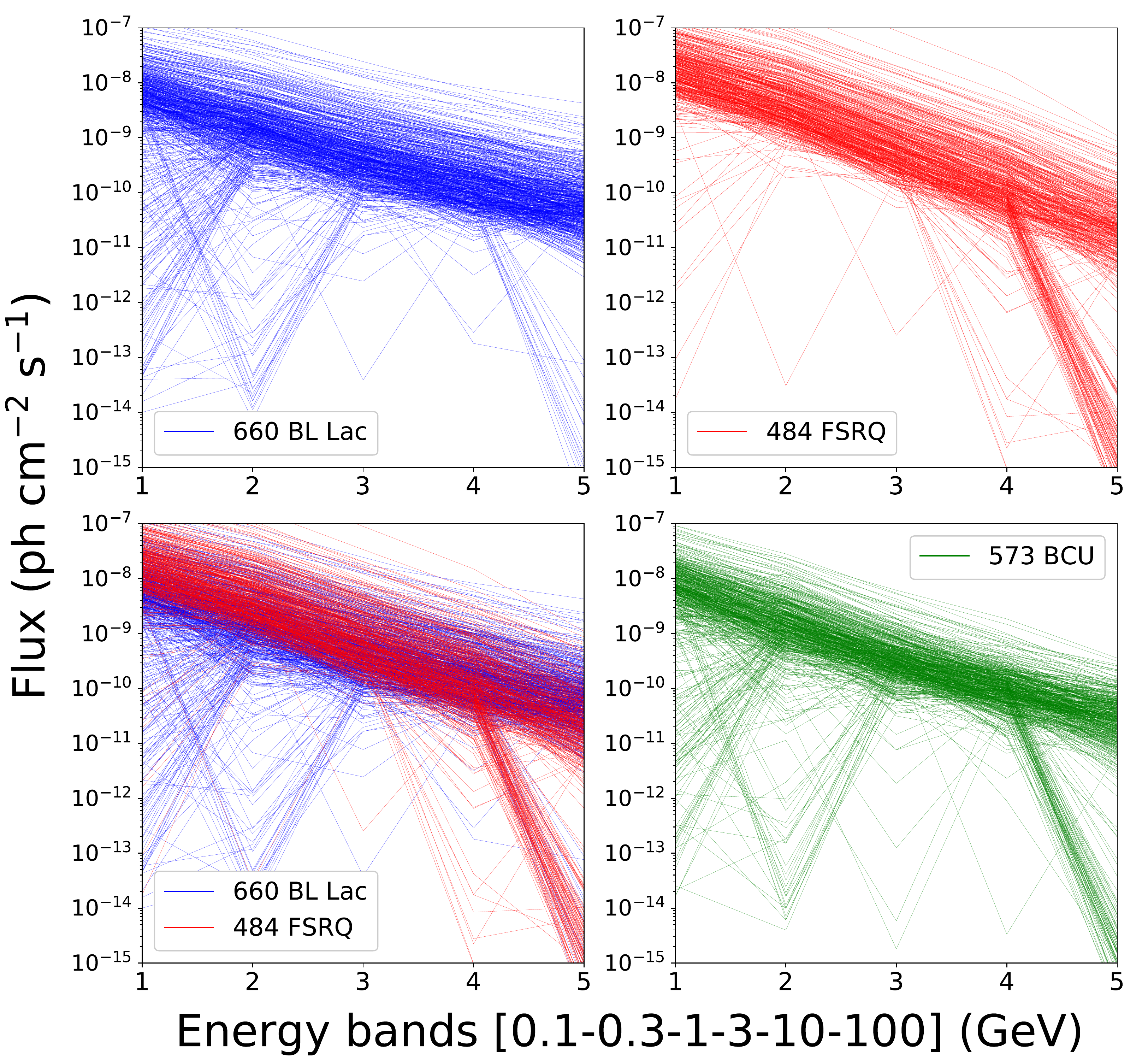}
\caption{Time integrated fluxes in 5 different energy bands. Band 1: 0.1--0.3 GeV; Band 2: 0.3--1 GeV; Band 3: 1--3 GeV; Band 4: 3--10 GeV; Band 5: 10--100 GeV. Each curve represents a single source. BL Lacs (blue) are in the top-left, FSRQs (red) in the top-right, both are in the lower-left and BCUs (green) are in the lower-right.}
\label{<eband>}
\end{center}
\end{figure}

In order to further improve performance we use spectral information in addition to flux-sorted light curves. In the 3FGL catalog there are time-integrated flux values in 5 different energy bands: 0.1--0.3, 0.3--1, 1--3, 3--10, 10--100 GeV (Fig.~\ref{<eband>}). This set of parameters contains information of average spectral index, spectral curvature, hardness and flux ratios, peak energy and others. \citet{hassan, lefau} used various spectral parameters obtained from fluxes in different energy bands in order to classify BCUs into BL Lacs and FSRQs, showing the value of using spectral information.

In the range of flux values $\sim$ 10$^{-10}$ ph cm$^{-2}$ s$^{-1}$ and energy bands from 0.1--0.3 up to 1--3 GeV, mostly BL Lacs are present, while for the energy band 10--100 GeV FSRQs are more numerous for the lower values of fluxes (around $\sim$ 10$^{-10}$ ph cm$^{-2}$ s$^{-1}$). The majority of BL Lacs and FSRQs have different slopes, which is in part a reflection of different average power-law indices. As in the case of sorted light curves, BCUs tend to behave more like BL Lacs, which would suggest that a larger number of them  could belong to that class.

\subsection{Radio and X-ray fluxes}
\label{<RadionX>}

Looking beyond $\gamma$-ray features of blazars, interesting information can be obtained from a multiwavelength study of the sources and particularly from X-ray and radio flux. In this study we tested the possibility to use those two parameters to improve the performance of the network. We did not consider any optical spectroscopy data because when considering uncertain blazars, optical spectra are very often not available or not sufficiently descriptive of the nature of the source.

\begin{figure}
\begin{center}
\includegraphics[width=.5\textwidth]{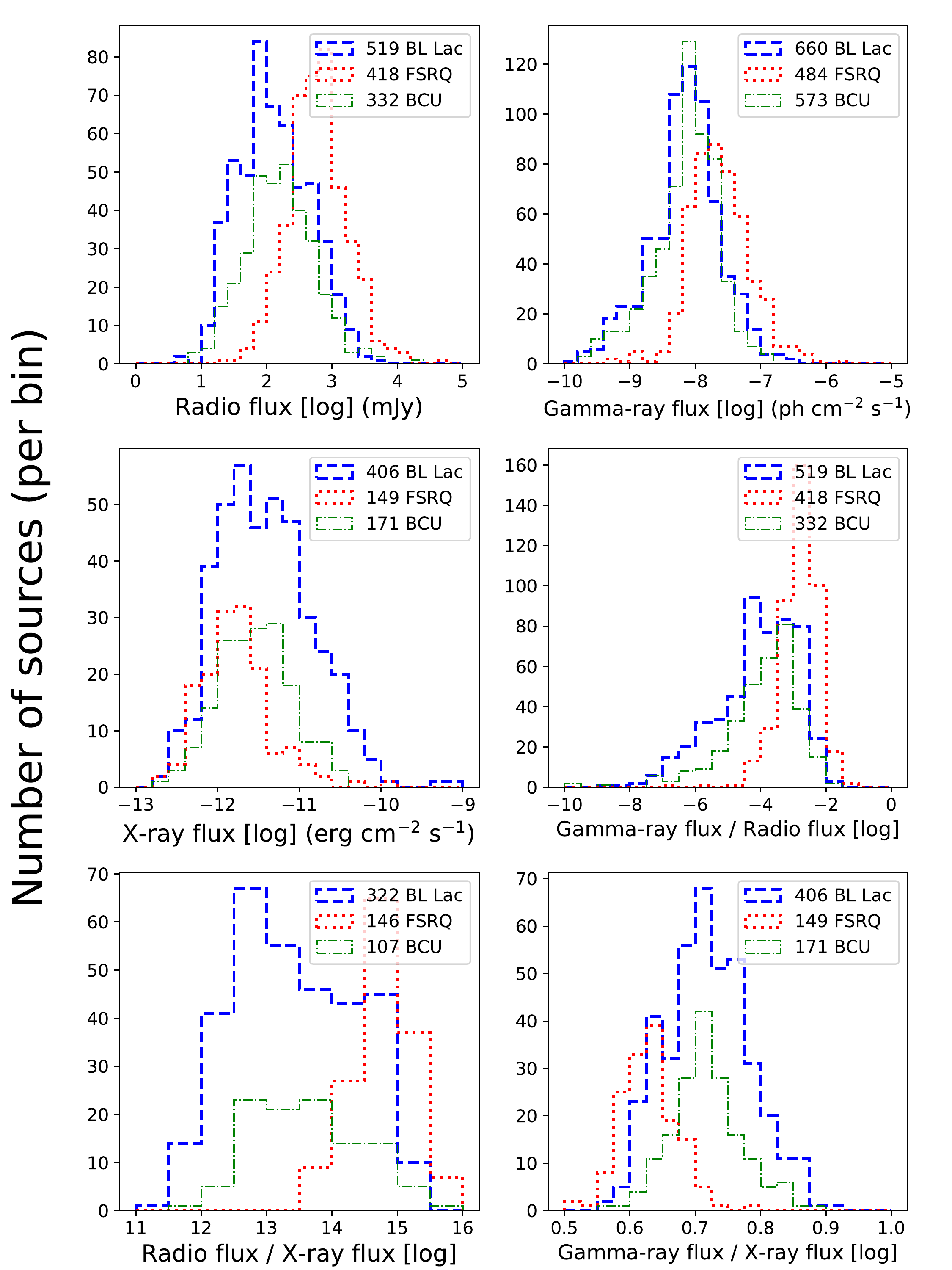}
\caption{
Top-left: 3FGL BL Lac, FSRQ and BCU radio flux histogram. Middle-left: X-ray flux histogram. Bottom-left: radio flux / X-ray flux histogram.
Top-right: 3FGL BL Lac, FSRQ and BCU gamma-ray flux histogram. Middle-right: Ratio of gamma-ray flux to radio flux. Bottom-right: Ratio of gamma-ray flux to X-ray flux.
\label{<radio>}}
\end{center}
\end{figure}

A particularly interesting parameter seems to be the ratio of radio (Sr) flux to the X-ray flux. In Fig.~\ref{<radio>} (3 plots on the left) the radio and X-ray flux histograms are shown. When the parameters are considered separately the contamination is not negligible (histogram on top-left and in the middle-left), but when the ratio Sr/X is considered it is possible to distinguish a clean area for BL Lacs where values are lower than $1\times10^{13}$ (13 in the plot). Unfortunately not all the known and uncertain blazars have both radio and X-ray flux data. However the final result is still appreciable because considering 3FGL blazars, 322 BL Lacs out of 660 have both radio and X-ray measurements and for 188 sources (28$\%$) the value of Sr/X ratio is lower than $4\times10^{13}$. Out of 484 FSRQs, 146 FSRQs have both radio and X-ray data and the value of Sr/X ratio is above $1\times10^{13}$ for all of them. Finally, out of 573 3FGL BCUs, 107 sources have both measurements while 57 show a value of the Sr/X ratio lower than $4\times10^{13}$. This means that the Sr/X ratio, although an overlap of data in higher values is not negligible, could be considered as an important area where a good separation for BL Lacs is possible.

For completeness we also consider time-integrated $\gamma$-ray flux (0.1--100 GeV band), and its ratios to radio and X-ray fluxes. The $\gamma$-ray flux was obtained by adding 5 time-integrated fluxes in 5 bands (Sect.~\ref{<spectrum>}). In Fig.~\ref{<radio>}, the 3 plots to the right show histograms of $\gamma$-ray flux (top-right), its ratio to radio (middle-right), and ratio to X-ray (lower-right). More on this topic will be discussed in the next section of the paper.

Radio and X-ray data were obtained from the the 3LAC catalog. Radio fluxes used were measured at frequencies of 1.4 GHz and 0.8 GHz; the X-ray fluxes were measured in the 0.1 -- 2.4 keV range \citep{3lac}.

Numbers on the plots in Fig.~\ref{<radio>} show how many sources have a given measurement and a flux value. In the case of ratios, numbers present how many sources have both given measurements and flux values.

\subsection{Data input}
\label{<DataInp>}

If parameter values (for an input neuron) vary over several orders of magnitude, it is common practice to use the logarithm of those values, in order to make it easier to find the right network settings to produce the optimal ANN function after training. 
Combinations with both original values and their logarithms were checked. The results are reported in Section ~\ref{<4.1>}.

The input data were normalized by subtracting their average value and dividing by their standard deviation so most of the input values fell between -1 and +1 for each input neuron.

The majority of sources do not have monthly $\gamma$-ray flux values above the detection threshold for all monthly time bins. Additionally the radio and X-ray data are missing for many sources. One way to deal with missing input data to a neural network is to set the inputs to zero. In this way, zero input acts as if there is no input neuron. Since the {\it Fermi}-LAT detector sweeps the sky continuously, the non-detection of $\gamma$-ray flux is due to low value of photon flux during the month and not observational constraints. The missing radio and X-ray data are due to low flux and/or observational constraints.

In our tests, the input radio and X-ray parameters for sources which do not have these values were set to zero. These zero values were not used in normalization and remained zero after normalization. The same is true if the logarithm of radio and X-ray flux was taken. 

The input data for missing monthly $\gamma$-ray flux is set to zero in the 3FGL catalog. These values were used in normalization after which they still had the lowest, but non-zero values. In case the logarithm was taken, these values were set to minimal value of monthly detected flux of $4\times10^{-11}$ cm$^{-2}$ s$^{-1}$ (of any source) before the logarithm was taken. After applying the logarithm, values were normalized as in the previous case.

\subsection{Network architecture}
\label{<net a>}

What follows is the architecture of the optimized network used in this work.

The number of neurons in input layer is equal to the number of input parameters: 48 neurons for 48 monthly $\gamma$-ray fluxes, 5 for time-integrated $\gamma$-ray fluxes in 5 different energy bands, 1 to 6 for any combination of radio, X-ray, integrated $\gamma$-ray flux, and their ratios. In total that is between 54 to 59 neurons in input layer.

The hidden layer consisted of 100 neurons. The choice was made by experimenting with single example for fixed number of training epochs. It was found that number of neurons should be higher than the number of input parameters (about 50 in our case) but after that the performance didn't change noticeably with further increase. The output layer consisted of two neurons. The activation function used in the hidden layer was hyperbolic tangent while for output neurons softmax (equation \ref{softmax}) function was applied which insured that sum of output neurons equals 1. In equation \ref{softmax} $\sigma_i$ is the output of the $i$-th neuron, $z_i$ is the input value to the activation function for the $i$-th neuron, $j$ is the summation index over neurons in the given layer.

\begin{align} \label{softmax}
\sigma_i =  \sum_{j} \dfrac{e^{z_i}}{e^{z_j}}.
\end{align}

The batch size was set to number of sources in training sample which insured smooth convergence. We tried three different Loss functions: mean squared error (used in \textit{C16}), mean absolute error (in order to reduce the impact of potential outliers or possibly wrongly labeled sources) and binary cross-entropy (equation \ref{bce}) which is typically used in binary classification. In equation \ref{bce}, $y_1$ and $y_2$ are desired values of 2 output neurons and can take values {0,1} or {1,0}, $p_1$ and $p_2$ are obtained values of 2 output neurons. Summation is performed over all sources in the batch sample.

\begin{align} \label{bce}
Loss = - \sum_{batch} ( y_1 \log p_1 + y_2 \log p_2 ).
\end{align}

The minimization algorithm used was adam-optimizer\footnote{\url{https://www.tensorflow.org/api_docs/python/tf/train/AdamOptimizer}}, a method for efficient stochastic optimization \citep{adamopt}, which converged quicker in our case than classical stochastic gradient descent.

\subsection{Training strategy}
\label{<Strat>}

Typically samples/sources for the training set and other sets are chosen randomly. The fluctuation in performance depending on which sources are taken might be important. In our case there are about 1000 labeled sources (BL Lacs and FSRQs), and it was found that the number of unclassified BCUs may vary significantly depending how training and other sets are chosen. In \citet{lefau} the same problem was noted, and we decided to test the strategy suggested by the authors, by training the network for many different training and testing samples and then selecting the set which is closest to the average results. We used 300 different train and test samples; the training set consisted of 70\%  and the test set of 30\% of the 3FGL classified blazars.

Aside from training the network on 300 different train and test samples, to avoid introduction of a second independent sample (with a yet smaller number of sources in it), an alternate strategy was used: the number of epochs was fixed for all combinations of input parameters and selections of training and testing samples, and the network was evaluated at the end; regularization was used to avoid over-fitting. The value for regularization was chosen so that it allowed the network to get close to the lowest test Loss function and to have it smoothly converge by the final epoch.

The desired outcome for training sample sources was set to  \{1,0\} and \{0,1\} for BL Lacs and FSRQs respectively. In this way the output neurons returned the likelihood of a source belonging to either class. Inputting parameters from known/labeled sources from the test sample into the trained network enables network evaluation. The two output neurons produce likelihood of a source being a BL Lac $L_{BL Lac}$ or  a FSRQ $L_{FSRQ}$ such that $L_{BL Lac} + L_{FSRQ} = 1$ for each source.

Network performance was evaluated by how many BCUs are left unclassified, applying a 90\% precision threshold \textit{C16}.

\section {Validation}
\label{<4>}

\subsection{Results}
\label{<4.1>}

The set of input parameters to the new network (Section ~\ref{<net a>}) were: a) 48 monthly $\gamma$-ray flux values sorted from lowest to highest (Section ~\ref{<variability>}), b) the 5 time integrated $\gamma$-ray flux values in 5 energy bands (Section ~\ref{<spectrum>}), c) any combination of the radio, X-ray, integrated $\gamma$-ray flux, and their ratios (Section ~\ref{<RadionX>}). Additionally some sources were excluded by applying the cut as described in Section ~\ref{<variability>}. 

All three types of Loss functions gave very similar results, however the best was obtained with mean absolute error. The application of the logarithm to any of the input parameters did not improve the final results, nor did it make them significantly worse.

Applying the new input parameters to the new network (48 monthly $\gamma$-ray sorted fluxes, 5 energy band fluxes, and just the radio to X-ray flux ratio), we improved the performance by decreasing the number of unclassified BCUs to 30 instead of 77 as reported in {\it{C16}}. When radio (Sr) and X-ray flux values were input separately instead only the Sr/X ratio, the number further falls to 15. This result is due to the fact that more BCUs have Sr and/or X values than both Sr and X. Out of a total of 573 BCUs in the 3FGL catalog, 332 have Sr values, 171 have X, and out of these 107 have both Sr and X (three left-hand plots in Fig.~\ref{<radio>}). Therefore, number of BCUs which have Sr and/or X is 396. If the ratio of Sr/X is added as additional input parameter to Sr and X alone, the performance remains the same. If integrated $\gamma$-ray flux and/or its ratios to Sr and X (three plots to the right in Fig.~\ref{<radio>}) are added as input parameters, the performance remains the same. These 3 parameters and Sr/X are combinations of parameters already used in the network and therefore they contain no true new information. All the number of unclassified BCUs mentioned above are average values of unclassified BCU from 300 different selections for training and testing samples.\\

\begin{figure}
\begin{center}
\includegraphics[width=.4\textwidth]{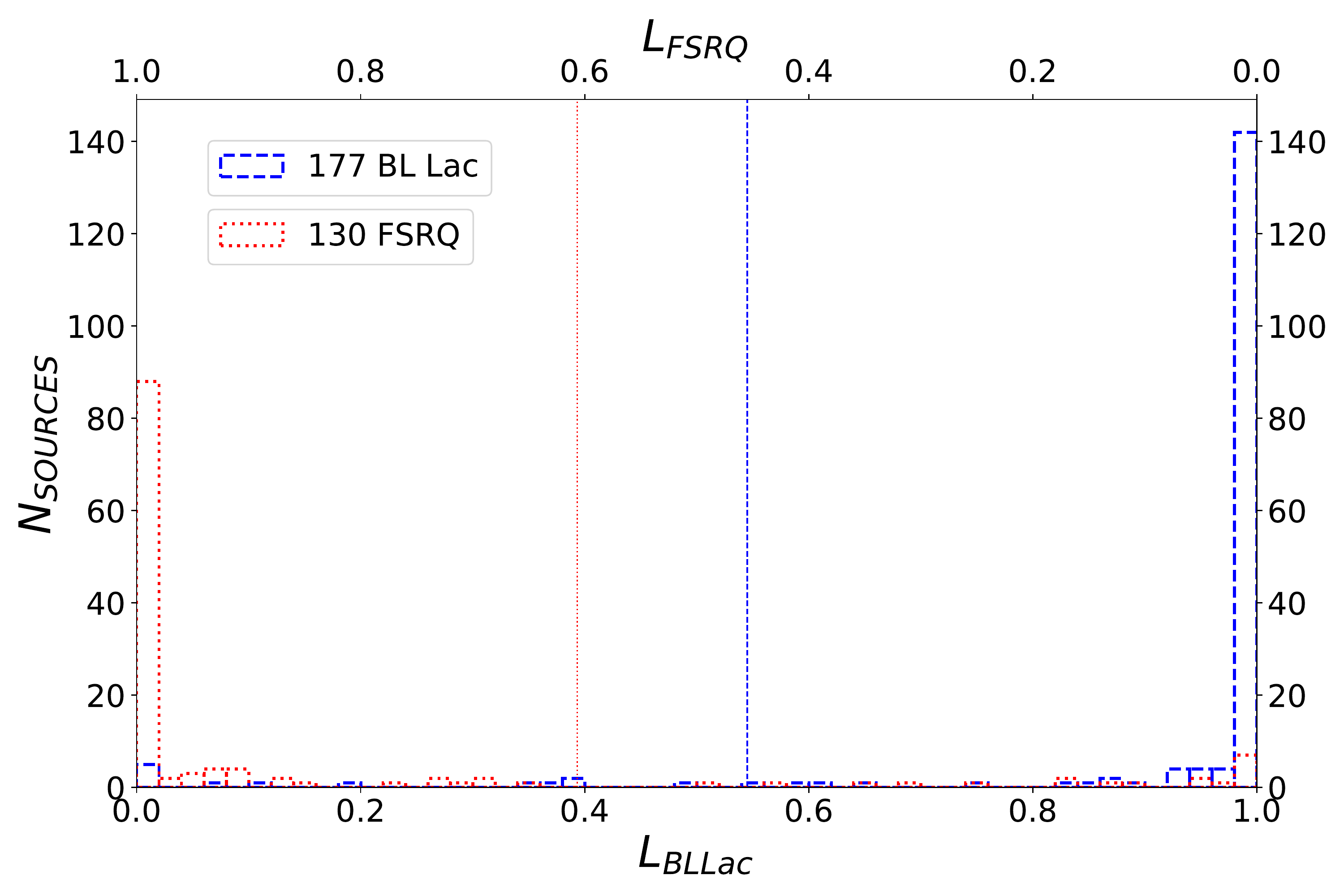}
\caption{Histogram of $L_{BL Lac}$ for BL Lacs and FSRQs from the test sample.  The blue (dashed) and red (dotted) vertical lines (at $L_{BL Lac} = 0.545$ and $L_{BL Lac} = 0.396$ respectively) present thresholds for BL Lacs and FSRQs such that precision of 90\% is obtained.} 
\label{<TestHist>}
\end{center}
\end{figure}

In Fig.~\ref{<TestHist>} the histogram of $L_{BL Lac}$ for BL Lacs and FSRQs from a representative test sample is presented. As expected BL Lacs concentrate towards $L_{BL Lac} \rightarrow 1$ while FSRQs $L_{BL Lac} \rightarrow 0$. The numbers of BL Lacs and FSRQs in test sample are 177 and 130 respectively. The sources for the test sample were chosen randomly but with two constraints: the numbers are 30\% of the total sample after application of the cut; the ratio of BL Lacs to FSRQs in the test sample (177:130 $\approx$ 1.36) was kept the same as the ratio in total sample (589:433 $\approx$ 1.36).

\begin{figure}
\begin{center}
\includegraphics[width=.4\textwidth]{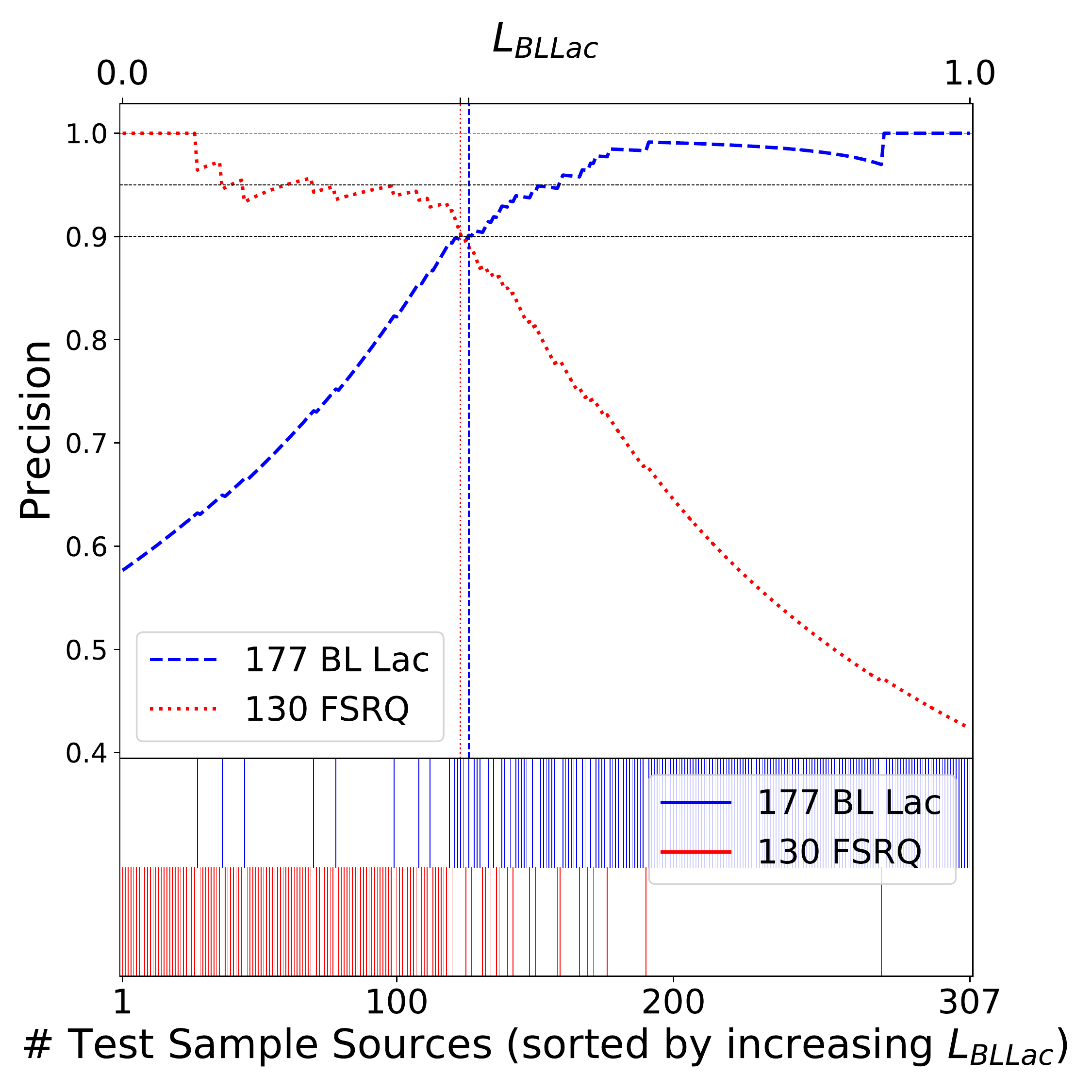}
\caption{Lower bar: BL Lacs (blue; upper lines) and FSRQs (red; lower lines) sources from the test sample sorted by increasing $L_{BL Lac}$ and at equal distance from each other. The $L_{BL Lac}$ does not increase linearly in the plot. The upper plot presents the change of precision with the $L_{BL Lac}$ threshold for BL Lacs and FSRQs. The threshold where precision reaches 0.9 for BL Lacs and FSRQs is $L_{BL Lac} = 0.545$ and $L_{BL Lac} = 0.396$ respectively (dashed blue and dotted red vertical lines). Precision is on average a monotonically increasing/decreasing function with $L_{BL Lac}$ for BL Lac/FSRQ. The zig-zag oscillations in precision is due to the finite and relatively small number of sources in the test sample.}
\label{<TestPrec>}
\end{center}
\end{figure}

The precision of the optimized neural network algorithm considering a threshold of 0.9 can be seen in Fig.~\ref{<TestPrec>}. Sources from the test sample are sorted by their $L_{BL Lac}$ (as in Fig.~\ref{<TestHist>}), but sources are at equal distance from each other and $L_{BL Lac}$ does not increase linearly. The threshold where precision reaches 0.9 for BL Lacs and FSRQs is $L_{BL Lac} = 0.545$ and $L_{BL Lac} = 0.396$ respectively (blue and red vertical lines). The threshold for BL Lacs is lower (more easily achieved) because BL Lacs are more numerous than FSRQs.

\begin{figure}
\begin{center}
\includegraphics[width=.4\textwidth]{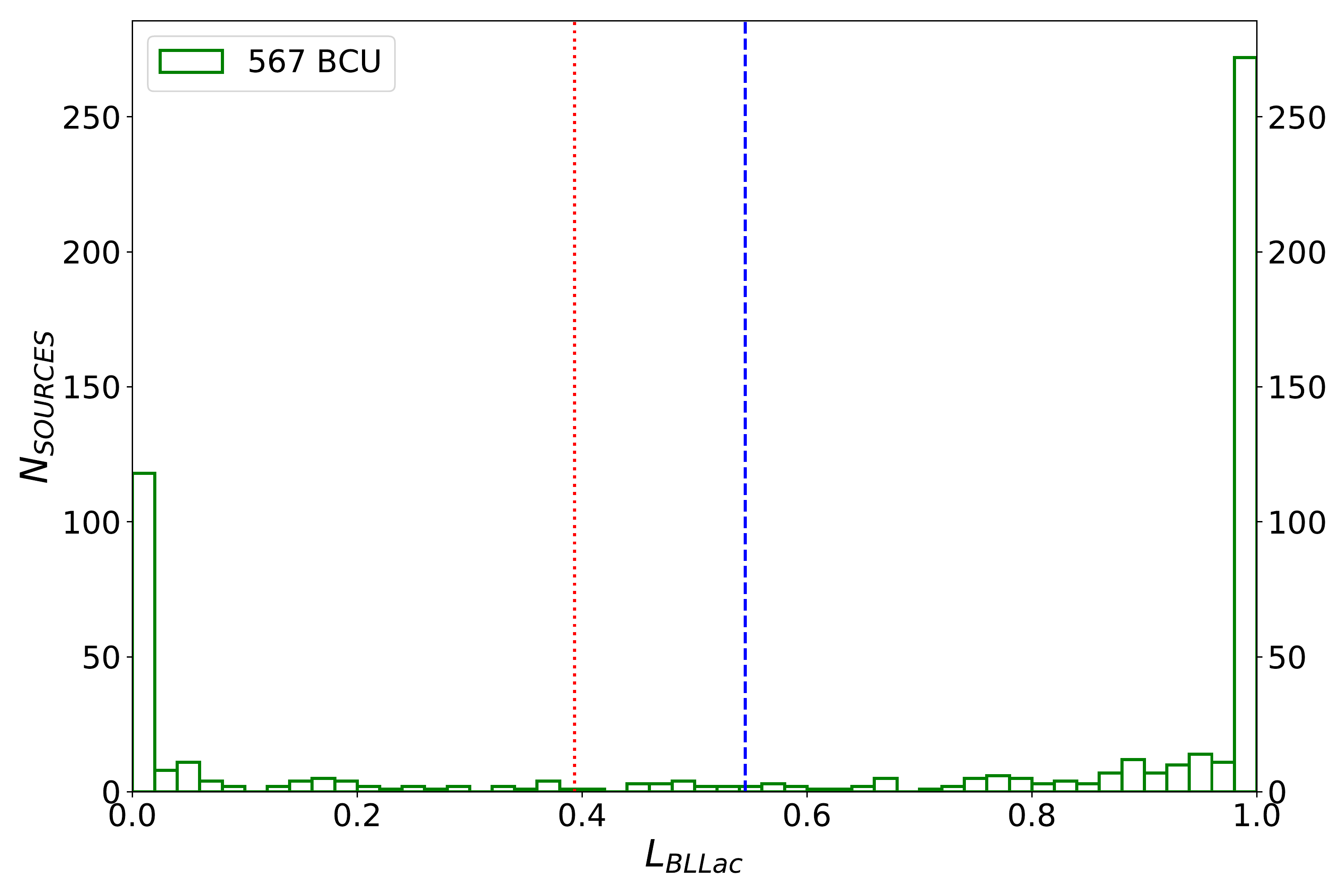}
\caption{Trained network applied to 567 BCU sources. Applying the threshold values of $L_{BL Lac} = 0.545$ and $L_{BL Lac} = 0.396$ (dashed blue and dotted red vertical lines) leaves 15 BCU unclassified, 378 classified as BL Lacs and 174 as FSRQs.} 
\label{<HistBCU>}
\end{center}
\end{figure}

Inputting BCU parameters into the trained network and applying the threshold values of $L_{BL Lac} = 0.545$ and $L_{BL Lac} = 0.396$ ( Fig.~\ref{<HistBCU>} ), as described above, the neural network leaves 15 BCUs unclassified ($0.396 > L_{BL Lac} > 0.545$), 378 classified as BL Lacs ($L_{BL Lac} > 0.545$) and 174 as FSRQs ($L_{BL Lac} < 0.396$). 

\subsection{Properties of classified BCUs}
\label{<4.2>}

In Table ~\ref{<tab1>} an example of 10 classified BCU sources is shown. The complete list of 567 classified BCUs is available in electronic format in which sources are sorted by increasing $L_{BL Lac}$. The classification is based on the 0.9 precision threshold obtained by comparing BCUs $L_{BL Lac}$ with dependence of precision on $L_{BL Lac}$ (Fig.~\ref{<TestPrec>}).  Note that BL Lac and FSRQ precision vs. $L_{BL Lac}$ are cumulative functions. Therefore a precision value listed for a source corresponds to average precision for all sources which have higher or lower $L_{BL Lac}$ than a given source.

\begin{figure}
\begin{center}
\includegraphics[width=.5\textwidth]{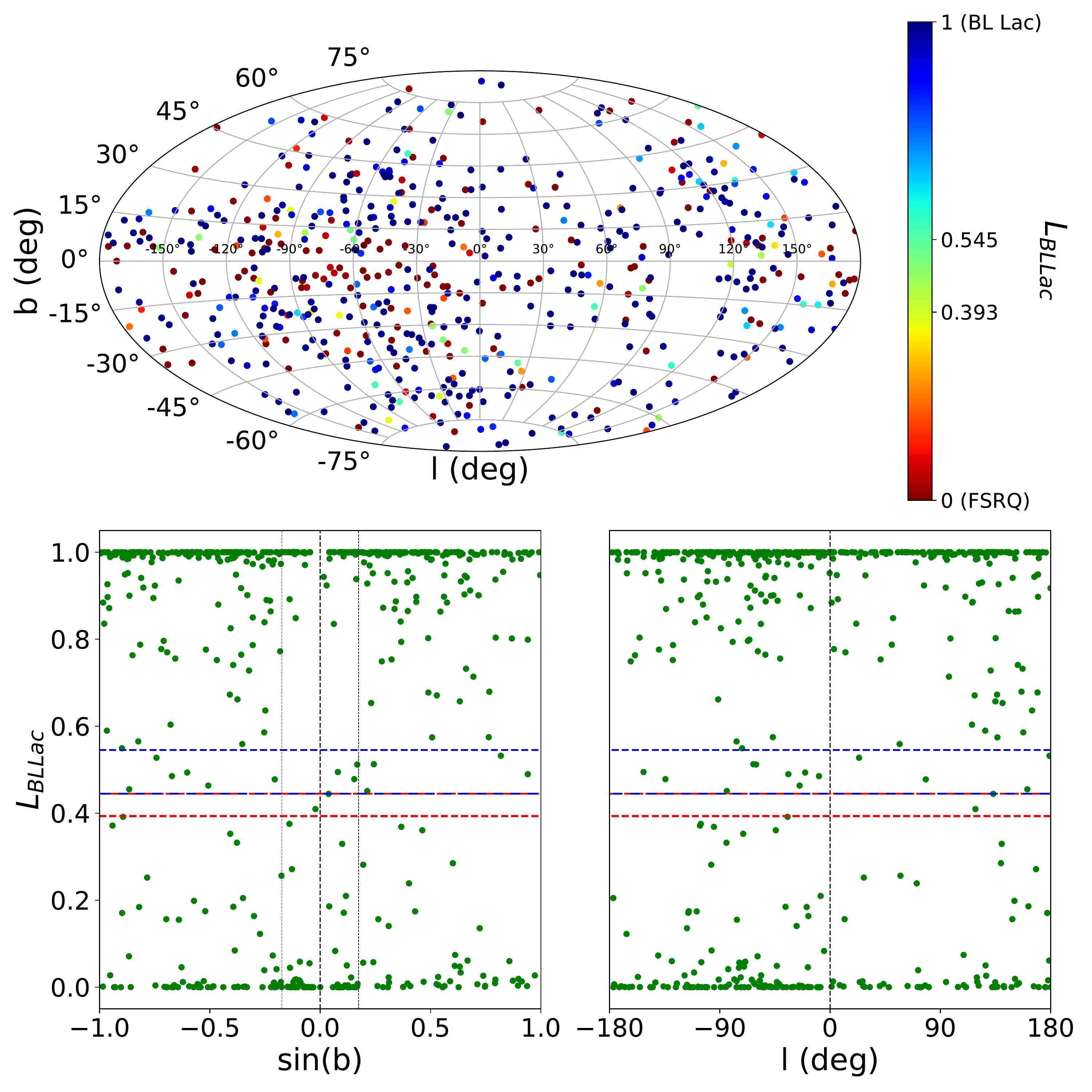}
\caption{Upper plot: Sky distribution in Galactic coordinates of 567 BCUs used in classification. Colors correspond to $L_{BL Lac}$. Thresholds of 0.545 and 0.396 for BL Lacs and FSRQs are shown in the color bar. Bottom plots:  $L_{BL Lac}$ of 567 BCUs vs. Galactic longitude (left) and latitude (right). Blue (upper) and red (lower) horizontal lines correspond to the two thresholds. Blue-red (middle) horizontal line at $L_{BL Lac}$ = 0.445 is a threshold corresponding to precision of about 89\% at which all BCUs can be classified. Bottom-left plot: two black dashed vertical lines around $b = 0^{\circ}$ correspond to $|b| = 10^{\circ}$.
} 
\label{<bcumap>}
\end{center}
\end{figure}

In Fig.~\ref{<bcumap>} the sky distribution in Galactic coordinates of 567 BCUs used in the classification is shown. The 15 BCUs that are left unclassified show no dependence on latitude or longitude, i.e. position with respect to the Galactic plane or Galactic center where $\gamma$-ray sources are more difficult to observe. The same can be noted for BCUs which are classified, but with less certainty ($L_{BL Lac}$ closer to threshold values). In order to quantify this we use a threshold $L_{BL Lac}$ = 0.445 corresponding to precision of about 89\% at which all BCUs can be classified. Then we use mean absolute error defined as $\sum |1 - L_{BL Lac}| / N$ for BL Lac candidates ($L_{BL Lac}$ > 0.445) and $\sum |0 - L_{BL Lac}| / N$ for FSRQs candidates ($L_{BL Lac}$ < 0.445). This quantity is an average measure of uncertainty of BCUs classification, i.e. how far away BCUs $L_{BL Lac}$ are from the peaks at $L_{BL Lac}$ = \{0,1\}. We found that this value is not bigger for BCUs at $|b| < 10^{\circ}$ than the ones at $|b| > 10^{\circ}$ meaning that BCUs near galactic plane are not classified with less certainty. The same result is obtained by using mean squared error.

\begin{table*}
\begin{center}
\caption{An example of 10 classified BCU sources is shown. The full list is available in electronic format. Columns: 3FGL name, Galactic latitude, Galactic longitude, $L_{BL Lac}$ (this work), precision value for BL Lac, precision value for FSRQ, BCU classification (this work), $L_{BL Lac}^*$ (\textit{C16}), BCU classification (\textit{C16}).
}
\label{<tab1>}
\begin{tabular}{lcccccccc}
\hline									
\hline									
Name & $b$ (deg) & $l$ (deg) & $L_{BL Lac}$ & $P_{BL Lac}$ & $P_{FSRQ}$ & Class & $L_{BL Lac}^*$ & Class$^*$  \\
\hline
3FGL J1532.7-1319 & 33.719 & 352.143 & 0.000 & & 1.000 & FSRQ & 0.000 & FSRQ \\
3FGL J1419.5-0836 & 48.376 & 336.849 & 0.003 & & 0.947 & FSRQ & 0.011 & FSRQ \\
3FGL J0939.2-1732 & 25.464 & 251.174 & 0.175 & & 0.929 & FSRQ & 0.155 & FSRQ \\
3FGL J2114.7+3130 & -11.884 & 77.994 & 0.482 & & & BCU & 0.514 & BCU \\
3FGL J0133.3+4324 & -18.815 & 130.957 & 0.731 & 0.909 & & BL Lac & 0.676 & BL Lac \\
3FGL J1344.5-3655 & 24.739 & 314.585 & 0.942 & 0.938 & & BL Lac & 0.825 & BL Lac \\
3FGL J2049.0-6801 & -35.896 & 326.659 & 0.995 & 0.977 & & BL Lac & 0.919 & BL Lac \\
3FGL J1434.6+6640 & 47.385 & 108.193 & 1.000 & 0.989 & & BL Lac & 0.995 & BL Lac \\
3FGL J0620.4+2644 & 5.632 & 185.708 & 1.000 & 0.986 & & BL Lac & 0.987 & BL Lac \\
3FGL J0649.6-3138 & -14.196 & 241.507 & 1.000 & 1.000 & & BL Lac & 0.978 & BL Lac \\
\hline
\end{tabular}
\end{center}
\end{table*}

\subsection{Comparison with \textit{C16} work}
\label{<4.3>}

The last two columns in Table ~\ref{<tab1>} present $L_{BL Lac}$ and BCU classification obtained in \textit{C16}.

\begin{figure}
\begin{center}
\includegraphics[width=.4\textwidth]{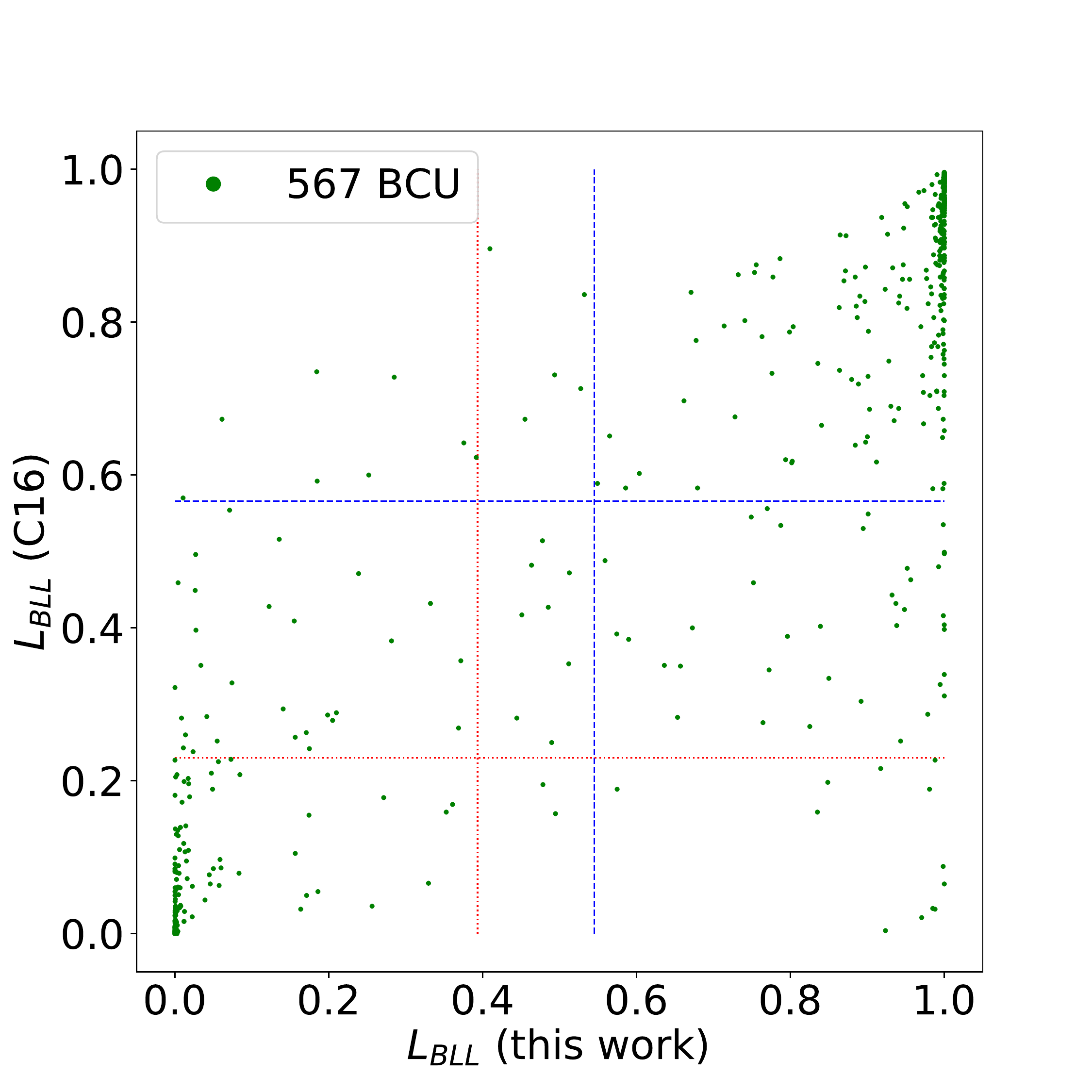}
\caption{567 BCUs used for classification in this work. Horizontal axis: $L_{BL Lac}$ obtained in this work. Vertical axis: $L_{BL Lac}$ obtained from the \textit{C16} paper. Blue (dashed) and red (dotted) lines present BL Lac and FSRQ 0.9 precision thresholds for new ($L_{BL Lac}$ = 0.545 and $L_{BL Lac}$ = 0.396) and \textit{C16} network ($L_{BL Lac}$ = 0.566 and $L_{BL Lac}$ = 0.230).} 
\label{<tfbf>}
\end{center}
\end{figure}

In Fig.~\ref{<tfbf>} 567 BCUs used for classification in this work are shown. The horizontal axis is $L_{BL Lac}$ obtained from this work while vertical is $L_{BL Lac}$ obtained from \textit{C16}. Blue and red lines present BL Lac and FSRQ 0.9 precision thresholds for new and \textit{C16} network. There is substantial overlap between the network classifications. 328 BCUs are classified as BL Lacs by both networks (upper-right area), 137 BCUs are classified as FSRQs by both networks (lower-left). Out of 75 unclassified BCUs (from 567 used in this work) by the \textit{C16} network, 38 are now classified as BL Lacs (middle-right) and 29 as FSRQs (middle-left) by the new network, while 8 (middle) are left unclassified by both networks. Out of 15 unclassified BCUs from the new network, 5 were classified as BL Lac (upper-middle) and 2 as FSRQs (lower-middle) by the \textit{C16} network, while 8 are unclassified by both as mentioned. Finally, 12 BCUs are classified as BL Lacs by new network and as FSRQs by the \textit{C16} one (lower-right); 8 BCUs are classified as FSRQs by new network and as BL Lacs by the \textit{C16} one (upper-left). In most cases, discrepancies between classifications are for sources which lie closer to threshold values with respect to majority of other sources, i.e. for sources which are classified with less certainty by both networks. Overall, it can be concluded that there is significant overlap between classifications, with the new network leaving fever BCUs left unclassified using the same precision threshold as in \textit{C16}. 

\section {Conclusion}
\label{<5>}

In this study, we explored the possibilities to increase the performance of a neural network method previously used for the classification of uncertain blazars.
We considered the improvement of performance by applying new input parameters (in the $\gamma$-ray, radio and X-ray range), and also from statistic adjustments. We developed an optimized version of the original algorithm improving the selecting performance of about 80$\%$. The final result of this study left 15 uncertain blazar sources instead of 77 in \textit{C16}. The rest of BCUs were classified into BL Lacs and FSRQs in the ratio of about two to one. This result confirms the machine learning techniques as powerful methods to classify uncertain astrophysics objects and particularly blazars. The artificial neural network technique could be a very worthwhile opportunity for the scientific community to select promising targets for multiwavelength rigorous classification and related studies at different energy ranges, mainly at very high energies by the present generation of Cherenkov telescopes and the forthcoming Cherenkov Telescope Array (CTA)\footnote{ \url{www.cta-observatory.org}} \citep{cta}.

We plan to use techniques described in this paper for classifying BCUs from the forthcoming 8-year LAT Source Catalog \textit{4FGL}. The new catalog will have more than 1000 BCUs obtained from twice as long observation period.

\section { Acknowledgments}                   

The authors would like to thank David J. Thompson (NASA Goddard Space Flight Center, Greenbelt, MD, USA) for review of the paper. For science analysis during the operation phase we acknowledge the {\it Fermi}-LAT collaboration for making the LAT results available in such a useful form. The authors also acknowledge their institutions for providing opportunity to carry out research. We would like to thank anonymous referee for discussion and suggestions leading to the improvement of this work.

\label{lastpage}
\end{document}